\documentclass[prb,aps,twocolumn]{revtex4}
\usepackage{epsfig,epsf,psfrag}
\usepackage{graphicx}
\usepackage{epic,eepic}
\usepackage{color,pstcol}
\begin{document}
\title{Self-consistent microscopic calculations for  non-local transport
through nanoscale superconductors}
\date{\today}

\author{R. M\'elin}

\affiliation{Institut NEEL,
  CNRS and Universit\'e Joseph Fourier, BP 166,
  F-38042 Grenoble Cedex 9, France}
\affiliation{Departamento de F\'{\i}sica Te\'orica de la Materia
Condensada C-V, Facultad de Ciencias, Universidad Aut\'onoma de Madrid,
E-28049 Madrid, Spain}

\author{F. S. Bergeret}
\affiliation{Departamento de F\'{\i}sica Te\'orica de la Materia
Condensada C-V, Facultad de Ciencias, Universidad Aut\'onoma de Madrid,
E-28049 Madrid, Spain}

\author{A. Levy Yeyati}
\affiliation{Departamento de F\'{\i}sica Te\'orica de la Materia
Condensada C-V, Facultad de Ciencias, Universidad Aut\'onoma de Madrid,
E-28049 Madrid, Spain}

\begin{abstract}
We implement self-consistent microscopic
calculations in order to describe out-of-equilibrium
non-local transport in normal
metal-superconductor-normal metal hybrid structures in the presence of a
magnetic field and for arbitrary interface transparencies.
A four terminal setup simulating usual experimental situations is
described by means of a  tight-binding model.
We present results for the self-consistent order parameter and
current profiles within  the sample.
These  profiles illustrate a crossover from
a quasi-equilibrium to a strong non-equilibrium situation when increasing
the interface transparencies and the applied voltages. 
We analyze in detail the behavior of the non-local conductance in
these two different regimes.
While in quasi-equilibrium conditions this can be expressed
as the difference between elastic cotunneling and crossed Andreev
transmission coefficients, in a general situation additional
contributions due to the voltage dependence of the self-consistent
order parameter have to be taken into account. 
The present results provide a first step towards a self-consistent
theory of non-local transport including non-equilibrium effects and 
describe qualitatively a recent experiment [Phys. Rev. Lett. {\bf 97}, 237003 (2006)].
\end{abstract}
\maketitle

\section{Introduction}
\begin{figure}[b]
\centerline{\includegraphics[width=7cm]{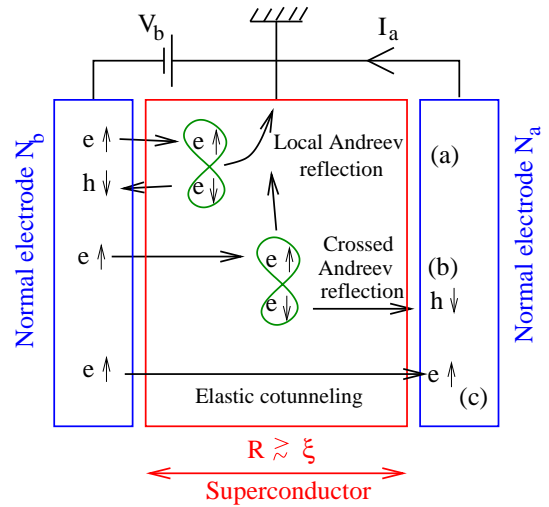}}
\caption{(Color online). 
Schematic representation of the electrical circuit used
in a non-local conductance experiment. The lowest order processes
are shown on the figure: local Andreev reflection~(a),
crossed Andreev reflection (CAR)~(b) and elastic cotunneling (EC)~(c).
\label{fig:carec}
}
\end{figure}

In addition to charge and spin,
future electronic devices may also
manipulate the non-local correlations
allowed by quantum mechanics, known as entanglement.
For instance a superconductor may be used as a source
of Einstein Podolsky Rosen pairs of electrons \cite{Choi,Martin}.
Local Andreev reflection \cite{Andreev} at a normal metal - superconductor
(NS) interface
is a process by which a spin-up electron incoming from the normal
side is reflected as a spin-down hole while a pair is transferred into
the superconductor:
pairs of electrons
penetrate the superconductor for an applied bias smaller than the gap
(see Fig.~\ref{fig:carec}a).
For an opposite applied voltage, the superconductor S emits correlated
pairs of electrons into the normal metal N.
Andreev reflection takes place in a coherence volume of size
$\xi_0$, the characteristic length associated to the superconducting
gap $\Delta$. Two separate normal electrodes connected to a superconductor
within a distance of order $\xi_0$ may thus be coupled by ``non-local''
or ``crossed''
Andreev processes $^{1,2,4-30}$ (CAR, see Fig.~\ref{fig:carec}b).
On the other hand another type of non-local process may take place. Electrons
can also tunnel across the superconductor from one normal electrode to the
other. This normal tunneling has been called
\cite{Falci} ``elastic cotunneling'' (EC) in analogy to 
similar processes taking place in
Coulomb blockaded quantum dots \cite{Nazarov}.

%%%%%%%%%%%%%%%%%%%%%%%%%%%%%%%%%%%%%%%%%%%%%%%%%%%%%%%%%%%%%%%%%%%%%%
%%%%%%%%%%%%%%%%%%%%%%%%%%%%%%%%%%%%%%%%%%%%%%%%%%%%%%%%%%%%%%%%%%%%%%

Motivated by the possibility of creation of  non-local coherent states by 
means of CAR processes,
several  experiments have been performed
recently\cite{Beckmann,Russo,Cadden} on  superconducting
structures connected to normal or ferromagnetic metallic electrodes. Depending on the type of S/N interfaces, geometry and range of parameters,   different
behaviors of the non-local conductance or resistance have been observed.

According to lowest order perturbation theory
\cite{Falci} in the tunnel amplitudes, EC and CAR have the same transmission
coefficient, once an average over the Fermi wave-length scale
or over disorder is carried out. Since an opposite charge is
transmitted by EC and CAR,
it is deduced that the non-local conductance vanishes in this limit.
A  finite non-local signal may be restored by higher order
tunneling processes\cite{Melin-Feinberg-PRB,Melin-Duhot-EPJB,Melin-wl} or by
Coulomb interactions \cite{Levy-Yeyati}.
Moreover,  out-of-equilibrium effects may play an important role on the
non-local transport as was suggested in Ref.~\onlinecite{Golubev}  for
a superconducting quantum dot 
and in Ref.~\onlinecite{Kalenkov}  for normal electrodes connected to
a three dimensional superconductor.

Experimental data on the other hand are
 still not well understood. In particular the experiment of Ref.~\onlinecite{Russo} on a planar NISIN structure
(where I stands for an insulating barrier) has provided
unexpected experimental evidence for a non-local signal
dominated either by EC  or by CAR  depending on  the value of the applied bias.
Also importantly,  this experiment 
\cite{Russo} has shown a suppression of the non-local signal when an external
magnetic field, much smaller than the critical one,  was applied. 
Other experiments 
on NSN hybrid structures provide evidence for
charge imbalance effects as the voltage approaches
the temperature-dependent gap value\cite{Cadden} and  in a temperature window
close to the superconducting critical temperature\cite{Beckmann}.

Although non-local transport in SN structures has been addressed theoretically 
by several works in recent years, none of these provide
a full self-consistent model for describing the superconducting order
parameter, the current profile and the non-local conductance  of  the S region, as well as the magnetic field dependence of these quantities.  

\begin{figure}[t]
\centerline{\includegraphics[scale=0.3]{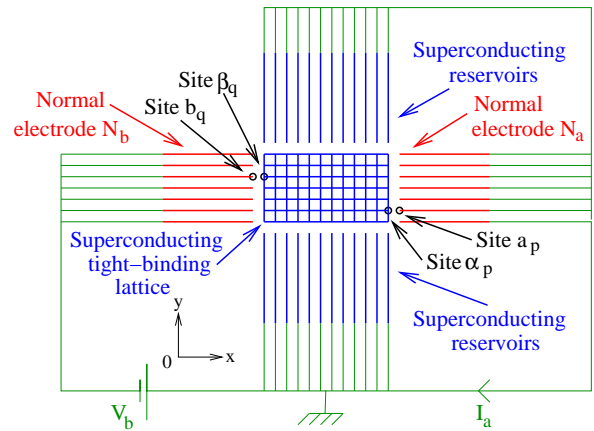}}
\caption{(Color online). Schematic representation of the tight-binding model
used in our simulations and of the electrical circuit (green lines).
The self-consistent gap is evaluated on a
 square lattice with $M=R_y/a_0$
transverse channels and a length $R_x=N a_0$,
with $a_0$ the lattice spacing.
Normal electrodes N$_b$ (N$_a$) are connected to the square lattice
at left (right). Superconducting reservoirs (with zero
superconducting phase) are connected on
top and bottom. The normal and superconducting electrodes
are modeled by a collection of one dimensional channels.
\label{fig:NINSS}
}
\end{figure}

To bridge this gap, we analyze here  a NISIN  planar structure
using a microscopic model which we solve
self-consistently in order to obtain the current profile inside the
superconductor in the presence of an applied voltage and magnetic field.
To adequately represent a non-local transport measurement setup
in the self-consistent calculation it is important to include additional
superconducting leads which allow to remove from the sample the injected
current (see Fig. \ref{fig:NINSS}). 
This implies a substantial difference with conventional ``two
terminal'' measurements, for which the  current is the same on both leads.
A two-terminal situation would be recovered by removing the
superconducting leads.

An important parameter  characterizing an actual experimental situation is
the value of the transparency for the barriers connecting the superconductor
with the normal electrodes. Although this parameter is difficult to be
controlled in experiments the available data correspond to very different
ranges as it was pointed out in
Ref.~\onlinecite{beckman2}. Within
our model it is possible to study non-local transport for arbitrary interface
transparency, applied bias and external magnetic field. This allows us to
analyze the crossover from a quasi-equilibrium situation at low barrier
transparency to the case of strong non-equilibrium for high transparency and
finite voltages. In agreement with previous works
\cite{Melin-Feinberg-PRB,Zaikin},
we find that increasing
barrier transparency leads to a dominance of EC over CAR transmission in the
non-local conductance. 
We also show that an applied magnetic field does not modify the balance
between EC and CAR.  The value of the EC and CAR transmission coefficients
increases upon application of voltage or magnetic field
(positive magneto-transmission).  
For simplicity we do not take into account in the present work the effect of
Coulomb interactions, although it has been realized that they can play an
important role for the case of low barrier transparencies
\cite{Levy-Yeyati,Duhot-Melin-PRB-Hikami}. We also restrict our analysis
to the ballistic case at zero temperature.

In addition to providing insight into the gap and current 
profiles in the superconducting region, our work demonstrates that a 
description of non-local transport in terms of linear relations involving 
the EC and CAR transmission coefficients only makes sense in a 
quasi-equilibrium situation when either the transparency of the S/N interfaces  or the bias voltage is low. 
If the superconductor is driven far from the equilibrium situation this 
description breaks down, and the resulting non-local conductance and 
resistance deviate considerably from their values in equilibrium. 
This case  corresponds to the experimental situation of
Ref.~\onlinecite{Cadden} 
where the non-local resistance was measured for currents close to the 
critical one. Within our model we are able to obtain the observed
change of sign of the non-local resistance as a function of the injected
current\cite{Cadden}. In particular we demonstrate that this change of sign is
related to the suppression of the local conductance and not due to the
dominance of CAR over EC processes.

%%%%%%%%%%%%%%%%%%%%%%%%%%%%%%%%%%%%%%%%%%%%%%%%%%%%%%%%%%%%%%%%%%%%%%
%%%%%%%%%%%%%%%%%%%%%%%%%%%%%%%%%%%%%%%%%%%%%%%%%%%%%%%%%%%%%%%%%%%%%%
The article is organized as follows. In the next
Sec.~\ref{sec:self-con-ef} we describe the  model
for a multiterminal NSN structure based on a tight-binding Hamiltonian with a
local pairing which will be determined self-consistently. We also give details
on how the electronic and transport properties of this model are obtained 
with the help of non-equilibrium Green functions.
In Sec. III we present results for the behavior of the complex order
parameter and  the current profiles in the presence of an applied magnetic
field, applied voltage and different values of the barrier transparency. 
We also analyze the behavior of the non-local conductance for both the quasi-equilibrium and non-equilibrium regimes.
Concluding remarks are provided in Sec.~\ref{sec:conclu}.

\section{Model for a multiterminal NSN structure}
\label{sec:self-con-ef}

We consider a superconducting region whose thickness is much smaller than the
coherence and the London penetration lengths. This region is connected
to four electron reservoirs as shown in Fig. 2.

We describe the central superconducting region by means of a tight-binding
model on a square lattice:
\begin{eqnarray}
\label{eq:H-BCS}
{\cal H}_{BCS}&=&-\sum_{\langle k,l\rangle,\sigma} \left(
t_{kl} c_{l,\sigma}^+ c_{k,\sigma}+
t_{l,k} c_{k,\sigma}^+ c_{l,\sigma}\right)\nonumber \\
&+& \sum_k \Delta_k \left( c_{k,\uparrow}^+ c_{k,\downarrow}
+c_{k,\downarrow}^+ c_{k,\uparrow}\right)\; .
\end{eqnarray}

The variable $\Delta_k$ in Eq.~(\ref{eq:H-BCS})
is the superconducting order parameter at site $k$ and a
summation over pairs of neighboring sites $\langle k,l\rangle$
is carried out in the kinetic term.
In order to take into account the effect of a magnetic field ${\bf B}={\bf rot
  A}$ one should introduce a phase in the hopping elements 
\begin{equation}
\label{eq:tkl}
t_{kl}=t \exp{\left(\frac{2i\pi}{\phi_0}\int_{{\bf r}_k}^{{\bf r}_l}
{\bf A}({\bf r}).{\bf dr}\right)}
,
\end{equation}
where ${\bf A}({\bf r})$ is the vector potential,
and $\phi_0=h/e$ is the flux quantum. In the following the magnetic field
will be measured in terms of $\phi/\phi_0$, i.e. the total flux through 
the central region in units of $\phi_0$.
Disorder could be  introduced in the form of an on-site random potential  
on each tight-binding site. 
We use the notations $R_x=N a_0$
and $R_y=M a_0$ for the dimensions of the lattice, where $a_0$ denotes the
lattice spacing.

The central  superconducting region 
is connected at the left to the normal electrode
N$_b$ and  at the right to the normal electrode N$_a$. 
In order to model a real experimental situation for non-local transport
we connect the  top 
and bottom  of the  central region to superconducting reservoirs
(see Fig.~\ref{fig:NINSS}) by highly transparent interfaces.
 Thus, a current injected through the left
interface can flow into these reservoirs. The leads, both the normal and
 superconducting, are described  by independent  one dimensional wires
 connected to each of the lateral sites of the central superconducting region.
 By varying the hopping terms $t_a$ and $t_b$ connecting the central 
region to the normal
 leads one can control the interface transparency $T_N$ between $0$ and
 $1$. The hopping parameter in the rest of the system is assumed to have 
the same value~$t$.

\subsection*{Calculational Methods}

The transport properties of this model can be conveniently expressed
in terms of the retarded $G^r$, advanced $G^a$ and
Keldysh $G^{+-}$ Green functions, which are matrices in the Nambu
space and depend on two site labels $k,l$.

The Nambu representation of the central region Hamiltonian 
takes the form
\begin{equation}
\label{eq:nambu-hopping}
\hat{h}_{kl}=
\left(\begin{array}{cc} 0 & \Delta_k \\
\Delta_k^{*} & 0 \end{array}\right)\; \delta_{kl} +
\left(\begin{array}{cc} t_{kl} & 0 \\
0 & -t_{kl}^{*} \end{array}\right) \; (1 - \delta_{kl}),
\end{equation}
where $t_{kl}$ 
is restricted to first neighbors and
was defined in Eq. (\ref{eq:tkl}).

The advanced and retarded Green functions are obtained from a recursive
algorithm. To this end we divide the central square lattice into layers 
along the $y$ direction which we label by an index $n$ ($1\le n\le N$).
We denote by $\hat{\bf G}^a_{n,n}$ the Nambu
advanced Green function projected on 
the sites corresponding to the $n$ layer. This is obtained as
\begin{eqnarray}
\label{eq:G1}
\hat{\textbf{G}}^a_{n,n}&=&
\left[\omega-\hat{\bf \Sigma}_n
-\hat{\textbf{T}}_{n,n-1}\hat{\textbf{\cal g}}^L_{n-1}\hat{\textbf{T}}_{n-1,n}
\right.\\
\nonumber
&&\left.
-\hat{\textbf{T}}_{n,n+1}\hat{\textbf{\cal g}}^R_{n+1}\hat{\textbf{T}}_{n+1,n}
\right]^{-1} \; ,
\end{eqnarray}
where the recursions take the form
\begin{eqnarray}
\label{eq:G2}
\hat{\textbf{\cal g}}^L_n&=&\left[\omega-\hat{\bf\Sigma}_n
-\hat{\textbf{T}}_{n,n-1}\hat{\textbf{\cal g}}^L_{n-1}\hat{\textbf{T}}_{n-1,n}
\right]^{-1}\\
\label{eq:G3}
\hat{\textbf{\cal g}}^R_n&=&\left[\omega-\hat{\bf\Sigma}_n
-\hat{\textbf{T}}_{n,n+1}\hat{\textbf{\cal g}}^R_{n+1}\hat{\textbf{T}}_{n+1,n}
\right]^{-1},
\end{eqnarray}
with the boundary conditions $\hat{\textbf{\cal g}}^L_0 = 
\hat{\textbf{\cal g}}^R_{N+1} = (i/t) \hat{\bf I}$, corresponding to 
the uncoupled normal wires in the wide band approximation. In the
recursive equations $(\hat{\bf \Sigma}_n)_{kl} = \hat{h}_{kl}
+ t \hat{\sigma}_z \hat{g}_S 
\hat{\sigma}_z \delta_{kl}(\delta_{k1}+\delta_{kM})$ with 
$\hat{g}_S = (-\omega \hat{I} + \Delta_0 \hat{\sigma}_x)/\sqrt{\Delta_0^2
- \omega^2}$, is the local
self-energy on the $n$ layer and $\hat{\bf T}_{n,n\pm1}$ contains the
hopping elements connecting neighboring layers. In the last expression
we have introduced the usual Pauli matrices in Nambu space $\hat{\sigma}_z$ 
and $\hat{\sigma}_x$.
The dependence on
energy $\omega$ is implicit in the Green functions in
Eqs.~(\ref{eq:G1})-(\ref{eq:G3}).
The Green function connecting two arbitrary layers $n$ and
$m$ can then be obtained from the relations
$\hat{\bf G}^a_{n,m} = \hat{\bf G}^a_{n,m-1} \hat{\bf T}_{m-1,m}
\hat{\bf g}^R_m$ or
$\hat{\bf G}^a_{n,m} = \hat{\bf G}^a_{n,m+1} \hat{\bf T}_{m+1,m}
\hat{\bf g}^L_m$.
In the absence of voltage applied on electrode N$_b$, the Keldysh
Green function of the superconductor takes the equilibrium form
\begin{equation}
\label{eq:Gpm_eq}
\hat{\bf G}^{+-,eq}_{n,n}(\omega)=n(\omega)
\left(\hat{\bf G}^a_{n,n}(\omega)-\hat{\bf G}^r_{n,n}(\omega)\right)
,
\end{equation}
where $n(\omega)$ is the Fermi distribution function
with zero chemical potential.
When a voltage $V_b$ is applied on electrode N$_b$
the Keldysh Green function of row $n$ is given by
\begin{eqnarray}
\label{eq:K}
\hat{\bf G}^{+-}_{n,n}(\omega) &=&
\hat{\bf G}^{+-,eq}_{n,n}(\omega)\\
&+&
\hat{\bf G}^{R}_{n,1}(\omega)\hat{\bf T}_{1,0}\delta \hat{\bf g}^{+-}_{0}
(\omega)\hat{\bf T}_{0,1}\hat{\bf G}^A_{1,n}(\omega)
\nonumber
,
\end{eqnarray}
where
\begin{equation}
\delta \hat{\bf g}^{+-}_{0}(\omega)=\frac{2i}{t} \left(
\begin{array}{cc} \delta n_e& 0 \\ 0 & \delta n_h \end{array}\right)
\hat{\bf I} \; , 
\end{equation}
with (assuming zero temperature)
\begin{eqnarray}
\delta n_e(\omega)&=&\theta(-\omega+eV_b)-\theta(-\omega)\\
\delta n_h(\omega)&=&\theta(-\omega-eV_b)-\theta(-\omega)
.
\end{eqnarray}

The pairing amplitude in the superconductor is set by the
anomalous component of the Keldysh Green function
given by Eq.~(\ref{eq:K}). Self-consistency equations 
at each site $k$ take the form
\begin{equation}
\label{eq:Deltanm}
\Delta_k =\lambda\int \frac{d\omega}{2i\pi} \left[\hat{G}^{+-}_{kk}
(\omega)\right]_{12} \; ,
\end{equation}
where $\lambda$ is the strength of the attractive electron-electron
interaction. This is chosen in order to obtain the same gap 
parameter $\Delta_0$ as in the upper and lower superconducting leads
in equilibrium conditions. 
Eq.~(\ref{eq:Deltanm}) is iterated until self-consistency
is achieved. The self-consistent gap, the current flow in the 
superconductor and the non-local conductance
are then evaluated. 
%%%%%%%%%%%%%%%%%%%%%%%%%%%%%%%%%%%%%%%%%%%%%%%%%%%%%%%%%%%%%%%%%%%%%%
\begin{figure}[t]
\centerline{\includegraphics[width=6cm]{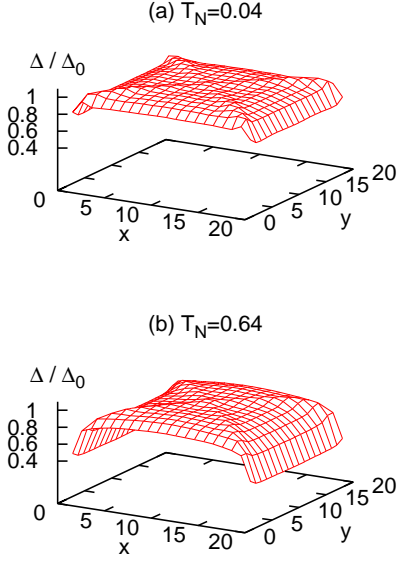}}
\caption{(Color online). 
Self-consistent gap profile  at zero voltage
for a system size $R_x/a_0=R_y/a_0=20$ and 
two different values  of the normal transparency.}
\label{fig:des_gap}
\end{figure}
%%%%%%%%%%%%%%%%%%%%%%%%%%%%%%%%%%%%%%%%%%%%%%%%%%%%%%%%%%%%%%%%%%%%%%
The current flowing from site $k$ to site $l$ is given by
\begin{equation}
\label{eq:Ikl}
I_{kl}=\frac{2e}{h} \int d\omega \left[
\hat{t}_{kl} \hat{G}^{+-}_{lk}(\omega)
-
\hat{t}_{lk} \hat{G}^{+-}_{kl}(\omega)
\right]_{11}
.
\end{equation}
A stringent test of self-consistency is provided by current conservation
at each site of the central region because current
is conserved only once self-consistency is verified\cite{yeyati95} .

%%%%%%%%%%%%%%%%%%%%%%%%%%%%%%%%%%%%%%%%%%%%%%%%%%%%%%%%%%%%%%%%%%%%%%
\begin{figure*}[t]
\centerline{\includegraphics[scale=.8]{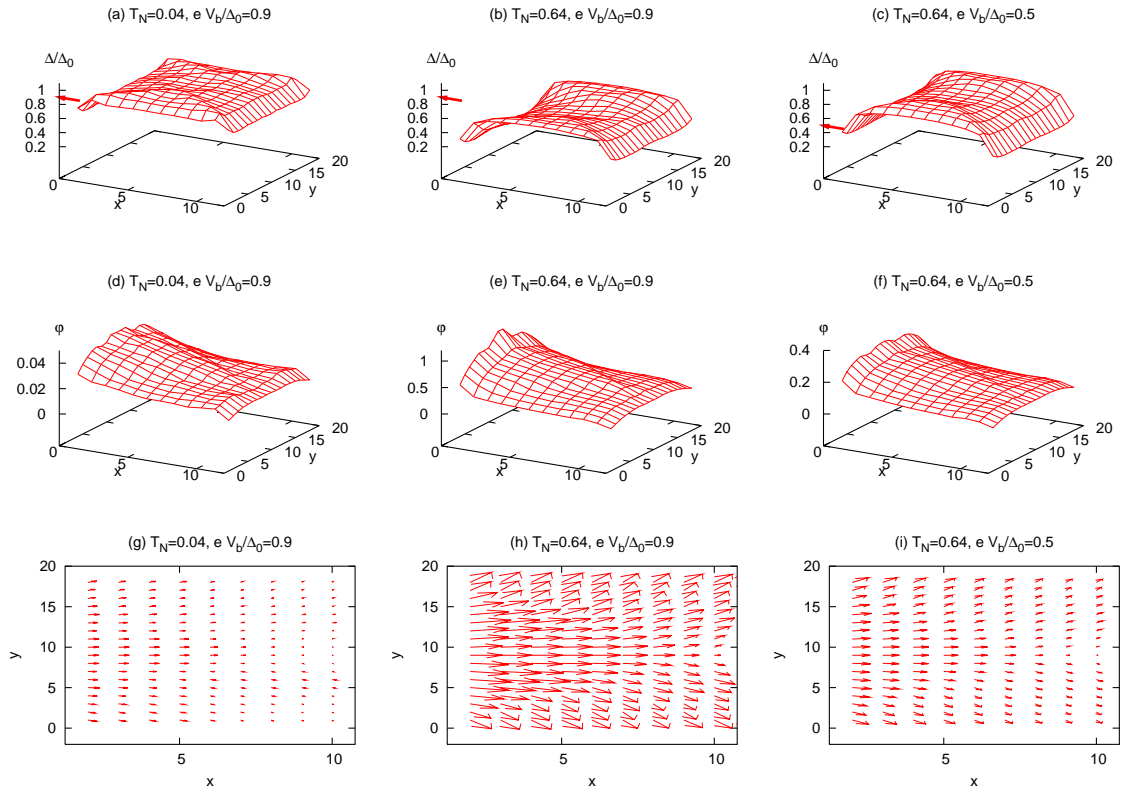}}
\caption{(Color online). 
Order parameter and current profiles for a sample size  $R_x/a_0=12$,
$R_y/a_0=20$ under an applied voltage $V_b$.
The upper panels show the profiles of the self-consistent gap.
The gap is normalized to its valued $\Delta_0$ in the superconducting
reservoirs. The voltage is indicated by an arrow on the vertical axis.
The middle panels show the phase profile and the lower ones  
the corresponding current maps.
The size of the arrows in the three lower panels is proportional to the value 
of the local current, with the same scaling factor for the three panels.
 The values of the parameters $T_N$ and $V_b$
are indicated on top of each panel.
\label{fig:gap-prof}
}
\end{figure*}
%%%%%%%%%%%%%%%%%%%%%%%%%%%%%%%%%%%%%%%%%%%%%%%%%%%%%%%%%%%%%%%%%%%%%%

%%%%%%%%%%%%%%%%%%%%%%%%%%%%%%%%%%%%%%%%%%%%%%%%%%%%%%%%%%%%%%%%%%%%%%
\begin{figure}
\centerline{\includegraphics[scale=0.7]{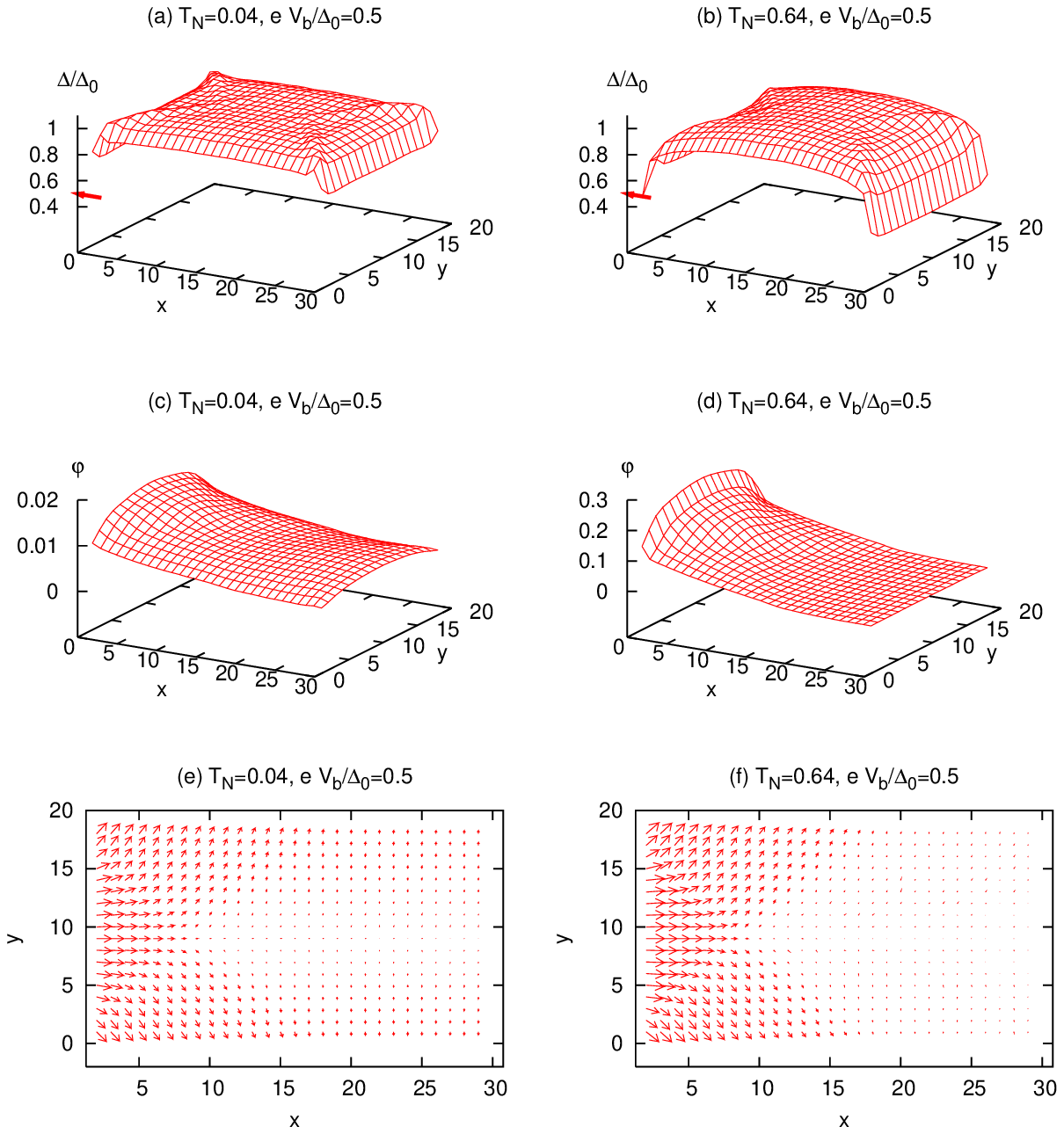}}
\caption{(Color online). Same as Fig. \ref{fig:gap-prof}
for sample dimensions $R_x/a_0=30$
and $R_y/a_0=20$. The scaling factor between current and size of arrow is 
not the same as in Figs.~\ref{fig:gap-prof}g, h and i, and it is different for
panels e and f.}
\label{fig:gap-prof2}
\end{figure}
%%%%%%%%%%%%%%%%%%%%%%%%%%%%%%%%%%%%%%%%%%%%%%%%%%%%%%%%%%%%%%%%%%%%%%

The non-local conductance can be computed
as ${\cal G}_{a,b}(V_a,V_b) = \partial I_a/\partial V_b(V_a,V_b)$,
where 
$I_a(V_a,V_b) = \sum_p I_{a_p \alpha_p}(V_a,V_b)$ is the total current flowing 
to electrode N$_a$ in response to voltages $V_a$ and $V_b$ on electrodes
a and b. We use here $V_a=0$, as in available experiments
\cite{Russo,Cadden,Beckmann}.
The notation $a_p$ ($b_q$) is used for 
the site at the right (left) interface formed by the superconductor and
the normal channel $p$ ($q$), while 
$\alpha_p$ ($\beta_q$) are used for the counterpart 
in the superconductor
(see Fig.~\ref{fig:NINSS}).  
In quasi-equilibrium conditions, i.e. when the variation of the
self-consistent order parameter with $V_b$ is negligible,
the non-local conductance
can be written as \cite{Melin-Feinberg-PRB}
\begin{equation}
\label{eq:Gab1}
{\cal G}_{a,b}(eV_b)=\frac{e^2}{h} \left(T_{CAR}(e V_b)
-T_{EC}(e V_b)\right)
,
\end{equation}
where the EC and CAR transmission coefficients 
in the present model are given by
\begin{widetext}
\begin{eqnarray}
\nonumber
T_{EC}(e V_b)&=& 2 t_a^2 t_b^2/t^2 
\sum_{p,q}   
\left(\left[\hat{G}_{\alpha_p,\beta_q}^a(e V_b)\right]_{11}
\left[\hat{G}_{\beta_q,\alpha_p}^r(e V_b) \right]_{11}
+ \left[\hat{G}_{\alpha_p,\beta_q}^a(e V_b)\right]_{22}
\left[\hat{G}_{\beta_q,\alpha_p}^r(e V_b)\right]_{22} \right)\\
\label{eq:TEC}
\\
\nonumber
T_{CAR}(e V_b)&=&2 t_a^2 t_b^2/t^2
\sum_{p,q} 
\left(\left[\hat{G}_{\alpha_p,\beta_q}^a(e V_b)\right]_{12}
\left[\hat{G}_{\beta_q,\alpha_p}^r(e V_b)\right]_{21}
+\left[\hat{G}_{\alpha_p,\beta_q}^a(e V_b)\right]_{21}
\left[\hat{G}_{\beta_q,\alpha_p}^r(e V_b)\right]_{12} \right) . \\
\label{eq:TCAR}
\end{eqnarray}
\end{widetext}

The sum over $p$ and $q$ in Eqs.~(\ref{eq:TEC}) and~(\ref{eq:TCAR})
corresponds to a sum over all one dimensional transverse channels
in the normal electrodes.
As mentioned above,
the parameters $t_a$ and $t_b$ are the hopping amplitudes
connecting electrodes N$_a$ and N$_b$ to the superconductor.
%%%%%%%%%%%%%%%%%%%%%%%%%%%%%%%%%%%%%%%%%%%%%%%%%%%%%%%%%%%%%%%%%%%%%%
\begin{figure}[t]
\centerline{\includegraphics[scale=.9]{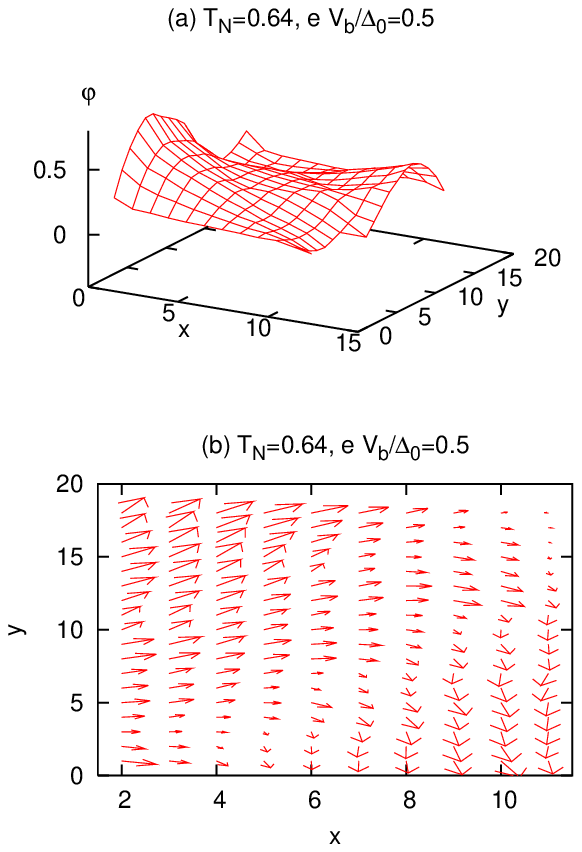}}
\caption{(Color online). 
(a) Surface map of the self-consistent phase under magnetic field, 
for sample dimensions $R_x/a_0=12$ and $R_y/a_0=20$.
The corresponding current map is shown on (b).
The flux is $\phi/\phi_0=1$.
The scaling factor between current and size of arrow is 
the same as in Figs.~\ref{fig:gap-prof}g, h and i. 
\label{fig:des_phase_field}
}
\end{figure}
%%%%%%%%%%%%%%%%%%%%%%%%%%%%%%%%%%%%%%%%%%%%%%%%%%%%%%%%%%%%%%%%%%%%%%

%%%%%%%%%%%%%%%%%%%%%%%%%%%%%%%%%%%%%%%%%%%%%%%%%%%%%%%%%%%%%%%%%%%%%%
\begin{figure*}[t]
\centerline{\includegraphics[scale=.7]{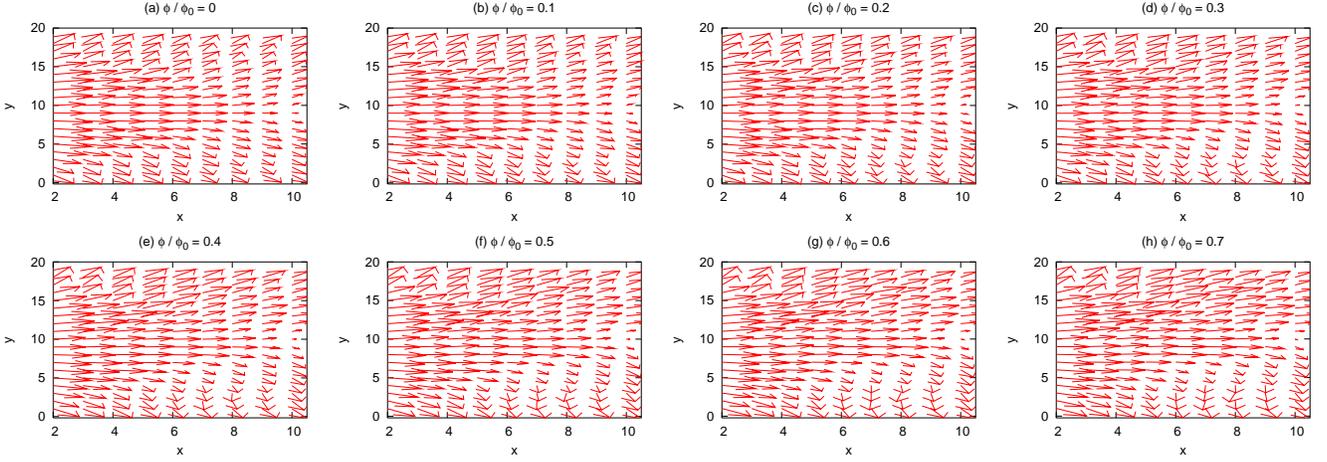}}
\caption{(Color online). 
Current map for increasing values of the magnetic flux $\phi/\phi_0$
and for $T_N=0.64$, $R_x/a_0=12$ and $R_y/a_0=20$.
The applied voltage $V_b$ is set to $0.9 \Delta_0$. An arrow with the same
length corresponds to the same current for all panels. 
The scaling factor between current and size of arrow is 
the same as in Figs.~\ref{fig:gap-prof}g, h and i. 
\label{fig:flux2}
}
\end{figure*}
%%%%%%%%%%%%%%%%%%%%%%%%%%%%%%%%%%%%%%%%%%%%%%%%%%%%%%%%%%%%%%%%%%%%%%

%%%%%%%%%%%%%%%%%%%%%%%%%%%%%%%%%%%%%%%%%%%%%%%%%%%%%%%%%%%%%%%%%%%%%%
\begin{figure*}
\centerline{\includegraphics[scale=.7]{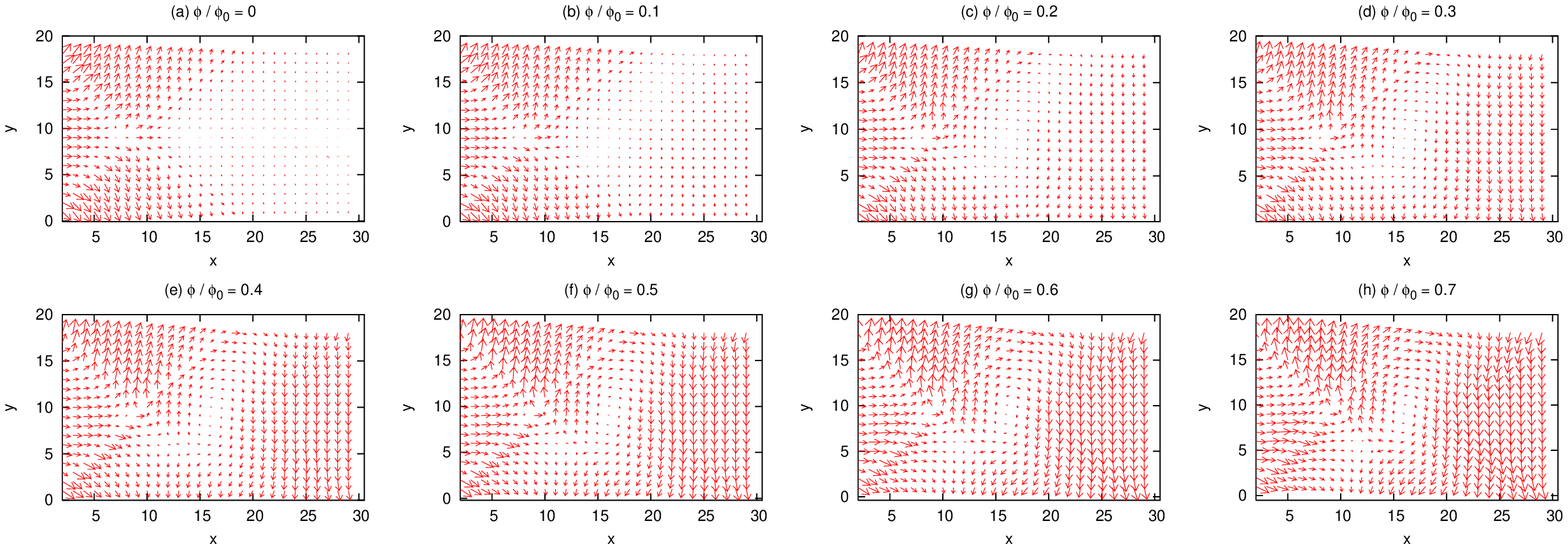}}
\caption{(Color online). 
Same as in Fig.~\ref{fig:flux2} for $T_N=0.64$, $eV_b=0.9\Delta_0$,
$R_x/a_0=30$ and $R_y/a_0=20$.
\label{fig:flux3}
}
\end{figure*}
%%%%%%%%%%%%%%%%%%%%%%%%%%%%%%%%%%%%%%%%%%%%%%%%%%%%%%%%%%%%%%%%%%%%%%
%%%%%%%%%%%%%%%%%%%%%%%%%%%%%%%%%%%%%%%%%%%%%%%%%%%%%%%%%%%%%%%%%%%%%%
\begin{figure}[t]
\centerline{\includegraphics[scale=.9]{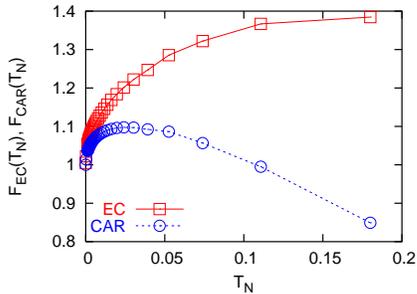}}
\caption{(Color online). 
Variation of the quantities $F_{EC}(T_N)$ and $F_{CAR}(T_N)$ defined in text,  as a function of the
normal transparency $T_N$.
A predominance
of EC over CAR is obtained for high values of interface transparencies.
The chosen system dimensions are $R_x=28 a_0$ and $R_y=20 a_0$.
\label{fig:des_vs_W}
}
\end{figure}
%%%%%%%%%%%%%%%%%%%%%%%%%%%%%%%%%%%%%%%%%%%%%%%%%%%%%%%%%%%%%%%%%%%%%%

%%%%%%%%%%%%%%%%%%%%%%%%%%%%%%%%%%%%%%%%%%%%%%%%%%%%%%%%%%%%%%%%%%%%%%
\begin{figure}[t]
\centerline{\includegraphics[scale=.9]{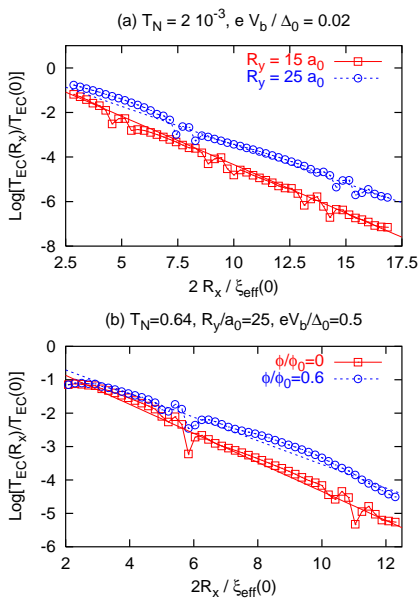}}
\caption{(Color online). 
Dependence of the EC transmission coefficient on the length
$R_x=Na_0$ of the superconductor. Panel (a) corresponds to low
transparency and two different values of the transverse dimension 
$R_y=15$ and $25 a_0$.
The sample dimension $R_x$ on panel (a) is normalized to 
the value $\xi_{eff}(0)$ of the coherence length for $R_y=15 a_0$.
Panel (b) corresponds to high transparency and two different
values of the magnetic flux. The sample dimension on panel (b) 
is normalized to the value $\xi_{eff}(0)$ of the coherence length 
for $\phi/\phi_0=0$. Out-of-equilibrium effects are negligible for
both panels (see insets of Fig.~\ref{fig:dep_voltage}).
\label{Rx_dependence}}
\end{figure}
%%%%%%%%%%%%%%%%%%%%%%%%%%%%%%%%%%%%%%%%%%%%%%%%%%%%%%%%%%%%%%%%%%%%%%
%%%%%%%%%%%%%%%%%%%%%%%%%%%%%%%%%%%%%%%%%%%%%%%%%%%%%%%%%%%%%%%%%%%%%%
\begin{figure}
\centerline{\includegraphics[scale=1.3]{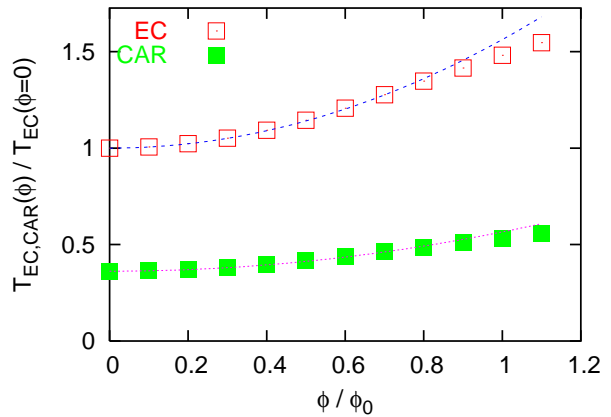}}
\caption{(Color online). Dependence of the transmission coefficients 
$T_{EC}(\phi)/T_{EC}(\phi=0)$,
and  $T_{CAR}(\phi)/T_{EC}(\phi=0)$ on the magnetic flux $\phi/\phi_0$.
We have chosen $T_N=0.64$, $R_x/a_0=20$ and $R_y/a_0=14$. The broken 
and dotted lines
indicate the quadratic fit of Eq.(\ref{eq:fluxdependence}).
\label{fig:dep_flux_mg}
}
\end{figure}
%%%%%%%%%%%%%%%%%%%%%%%%%%%%%%%%%%%%%%%%%%%%%%%%%%%%%%%%%%%%%%%%%%%%%%

%%%%%%%%%%%%%%%%%%%%%%%%%%%%%%%%%%%%%%%%%%%%%%%%%%%%%%%%%%%%%%%%%%%%%%
\begin{figure*}[t]
\centerline{\includegraphics[scale=.95]{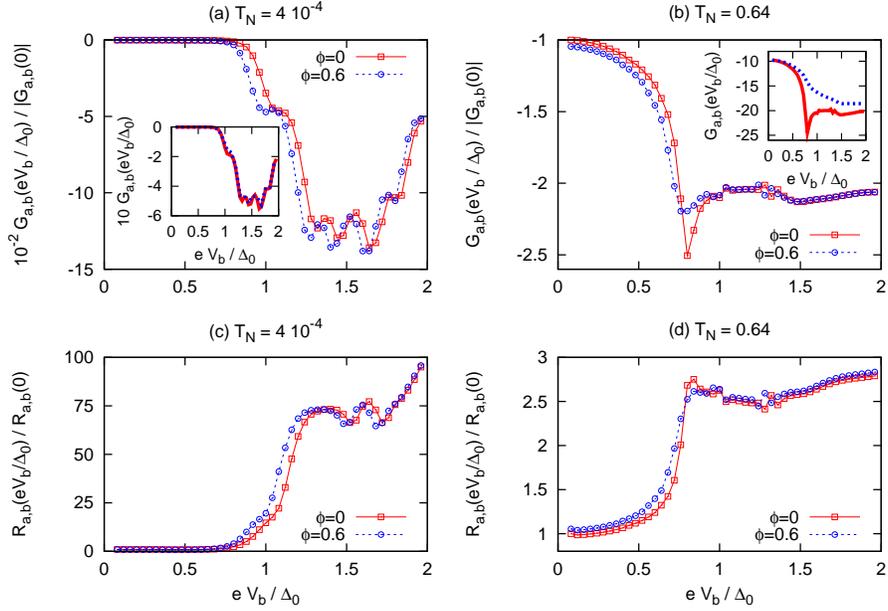}}
\caption{(Color online). 
Voltage dependence of the non-local conductance
(upper panels), and of non-local resistance (lower panels),
normalized to their value at zero bias in zero applied
magnetic field. The values of $T_N$ and $\phi$ are indicated on each panel.
The insets of the upper panels illustrate the comparison between
the full calculation ${\cal G}_{a,b} = \partial I_a/\partial V_b$ (solid line)
and the expression approximation ${\cal G}_{a,b} = e^2/h(T_{CAR}-T_{EC})$
with the transmission coefficients defined by Eqs.~(\ref{eq:TEC})
and (\ref{eq:TCAR}) (dashed line). The non local conductance
${\cal G}_{a,b}$ in the insets is given in units of $e^2/h$.
\label{fig:dep_voltage}
}
\end{figure*}
%%%%%%%%%%%%%%%%%%%%%%%%%%%%%%%%%%%%%%%%%%%%%%%%%%%%%%%%%%%%%%%%%%%%%%

%%%%%%%%%%%%%%%%%%%%%%%%%%%%%%%%%%%%%%%%%%%%%%%%%%%%%%%%%%%%%%%%%%%%%%
\begin{figure*}[t]
\centerline{\includegraphics[scale=.9]{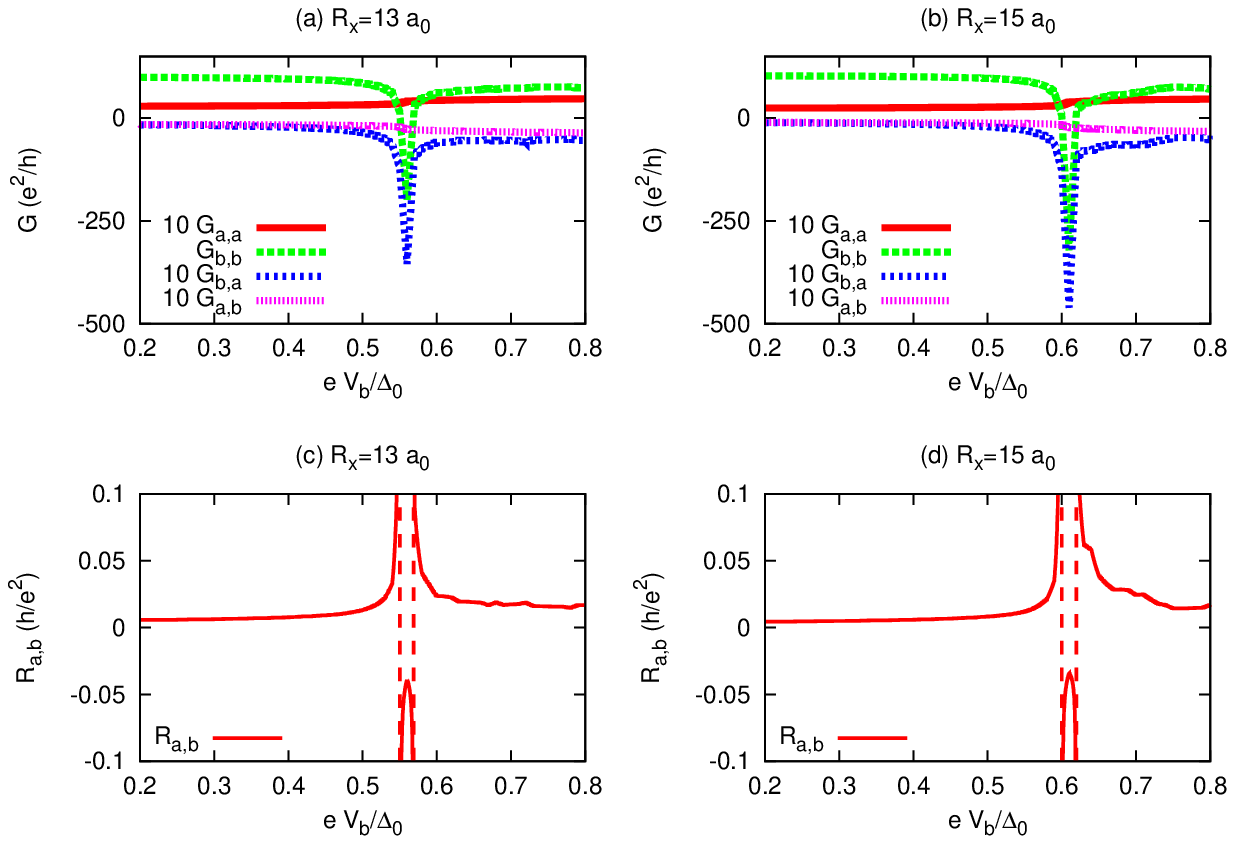}}
\caption{(Color online). 
Voltage dependence of the local and non-local conductances (upper
panels) and of the non-local resistance (lower panels)
for an asymmetric junction with $T_N=1$ at the $b$ interface and $T_N=0.1$
at the $a$ side, for $R_y=40 a_0$ and two values of $R_x$:
$13 a_0$ [panels (a) and (b)] and $15 a_0$ [panels (c) and (d)].
\label{fig:desI}
}
\end{figure*}
%%%%%%%%%%%%%%%%%%%%%%%%%%%%%%%%%%%%%%%%%%%%%%%%%%%%%%%%%%%%%%%%%%%%%%

\section{Results}
\label{sec:results}
\subsection{Self-consistent gap, phase and current profiles}

As a first step it is instructive to analyze the behavior 
of the complex order parameter which is obtained from the self-consistency
Eq. (\ref{eq:Deltanm}).  
In order to be able to accurately describe the spatial variations 
with a reasonable computational cost we fix  
$\Delta_0/t=0.1$, which roughly corresponds to a coherence 
length $\xi_0 \simeq 7 a_0$. The system sizes that we consider in this
work correspond typically to $R_x,R_y\sim 12-30\,a_0$ but in some
cases we use $R_x$ up to $60 a_0$ and $R_y$ up to
$40 a_0$.

The profile of the self-consistent order parameter amplitude (gap)
at zero voltage and zero magnetic field is shown on Fig.~\ref{fig:des_gap}
for a low ($T_N=0.04$) and 
a high ($T_N=0.64$) values of the interface transparencies.
As can be observed 
the gap is reduced at the contacts with the normal leads due to the
inverse proximity effect. This gap suppression becomes stronger as
the interface transparency increases. There is also a slight suppression
on the contacts with the upper and lower superconducting reservoirs
which appears due to its deviation from an ideal interface.
On the other hand, in equilibrium 
and for sufficiently large system size 
the gap in the middle of the central superconducting region
is close to its value in the reservoirs. One can also notice that the
gap profiles exhibit small ripples along certain directions, which in the
case of a square region ($R_x=R_y$)  correspond to the diagonal lines. 
These ripples reflect the ballistic character of our model 
which leads to constructive interference along semi-classical 
trajectories (for a discussion see Appendix).

An applied voltage on the left electrode N$_b$
leads to several modifications in the order 
parameter profiles. This is illustrated in Figs. \ref{fig:gap-prof}
and \ref{fig:gap-prof2}
for system sizes $R_x=12 a_0$ and $R_x=30 a_0$.
First, there is a reduction in the average gap value due to depairing effects.
This reduction is more pronounced near the contact where the current is
injected.
It is interesting to note that for the short system 
and high transparency [panel (b) in Fig. \ref{fig:gap-prof}]
the average gap becomes smaller than the applied voltage 
for $V_b/\Delta_0=0.9$.
In contrast for samples with $R_x$ sufficiently large compared to the
coherence length, the overall gap profile is
less sensitive to the applied voltage even for high transparency
[panel (b) in Fig. \ref{fig:gap-prof2}].

Secondly, the application of a finite voltage $V_b$ leads to a smooth
drop of the phase of the order parameter along the $x$ 
direction [panels (d)-(f) in Fig. \ref{fig:gap-prof}
and (c)-(d) in Fig. \ref{fig:gap-prof2}]. As can be observed
this is accompanied by a phase gradient in the $y$ direction
which changes sign at the midpoint $y=R_y/2$.

The properties of the self-consistent solution can be understood more
physically by analyzing the current profiles.  
These are shown in the lower panels of Figs. \ref{fig:gap-prof}
and \ref{fig:gap-prof2}.
As a general remark, the profiles illustrate how the injected current
from the left electrode gradually leaks to the upper and lower
superconducting electrodes. As expected, the amount of
current reaching the right normal lead is controlled
by the barrier transparency, the applied voltage and the system size. 
For sufficiently large system size one can clearly appreciate an
exponential suppression of the injected current on the $\xi_0$ scale
[panels (e) and (f) in Fig. \ref{fig:gap-prof2}]. In addition
the current profiles in the case of the larger system exhibit
some structure along the lines $y=x$ and $y=R_y-x$ which can be
associated with the already mentioned ripples appearing in the gap
profiles.

In the presence of applied voltage and magnetic field the phase
profile becomes more complex with a modulation both in the $x$
and $y$ directions. As shown in Fig. \ref{fig:des_phase_field} this
is associated with the appearance of currents along the $y$
direction which tend to screen the applied field. These currents
are essentially superimposed to the profile arising
from the injected current through the left electrode.
On the other hand, for this range of parameters the effect of the 
magnetic field on the amplitude of the order parameter corresponds to a 
uniform reduction (not shown here).

The superposition of injected and screening currents can be clearly
observed in the evolution of the current profiles for increasing values 
of the magnetic field, shown in Figs.~\ref{fig:flux2} and~\ref{fig:flux3}
for the cases $R_x=12 a_0$ and $R_x=30 a_0$ respectively. These plots
correspond to the case of high transparency ($T_N=0.64$) and
high voltage ($V_b=0.9\Delta_0$) for which the injected and screening
currents have a similar size when $\phi/\phi_0 \simeq 1$.
For a given total flux $\phi$ the screening currents are, however,
smaller for the case of the shorter system due to the larger gap
suppression induced by size effects and by the applied voltage.

\subsection{Low bias non-local conductance}
\label{sec:toto}
In this subsection we analyze the non-local transport properties
in the regime $eV_b \ll \Delta_0$.
As discussed in the section on calculational methods the non-local
conductance in quasi-equilibrium conditions
can be decomposed into EC and CAR contributions, given
by Eqs. (\ref{eq:TEC}) and (\ref{eq:TCAR}). 
These two contributions cannot be disentangled in experiments, but
theoretically it is  
convenient to analyze them in a separate way,
especially in the case of low transparency tunnel barriers
where the two contributions nearly cancel in the non-local
conductance\cite{Falci}.

The variation of the low bias EC and CAR transmission coefficients with
barrier transparency $T_N$ is shown on Fig.~\ref{fig:des_vs_W}.
We find it convenient to represent in this figure 
the quantities $F_{EC}(T_N)$ and $F_{CAR}(T_N)$ defined as
\begin{eqnarray}
\label{eq:FEC}
F_{EC}(T_N)&=&\frac{(T_N)^{-2} T_{EC}(T_N)}
{\lim_{T_N\rightarrow 0} (T_N)^{-2} T_{EC}(T_N)}\\
\label{eq:FCAR}
F_{CAR}(T_N)&=&\frac{(T_N)^{-2} T_{CAR}(T_N)}
{\lim_{T_N\rightarrow 0} (T_N)^{-2} T_{EC}(T_N)}
.
\end{eqnarray}

This normalization allows to analyze the variation of
$T_{EC}$ and $T_{CAR}$ beyond the dominant $\sim T_N^2$ dependence
which is common to both coefficients.
As can be observed, transmission of
electrons dominates over crossed Andreev processes for highly
transparent interfaces while they become almost equal
in the $T_N \rightarrow 0$ limit. This is an effect previously
discussed in literature\cite{Falci,Melin-Feinberg-PRB}
which is not modified significantly in our self-consistent
calculation.

It is also interesting to analyze the dependence of the
EC and CAR coefficients on the superconductor length $R_x$.
This is illustrated in Fig.~\ref{Rx_dependence}. We choose
to represent in this figure only $T_{EC}$ as $T_{CAR}$ exhibits
the same distance behavior. We find that except for tiny fluctuations
at certain values of $R_x$ the normalized transmission coefficients 
follow an exponential decay with sample dimensions of the type
$T(R_x) \sim \exp{\left(-2 R_x/\xi_{eff}\right)}$, where $\xi_{eff}$
is an effective coherence length which is sensitive to the
applied voltage and magnetic field.
As shown in panel (a) of Fig.~\ref{Rx_dependence}, the effective
coherence length is also sensitive to the transverse dimension
$R_y$, increasing slightly as $R_y$ increases. 
From Fig.~\ref{Rx_dependence}a we can estimate the values of $\xi_{eff}$. The 
coherence length for $R_y = 15a_0$ is $\xi_{eff}\sim 7 a_0$, while for 
$R_y = 25a_0$ is $\xi_{eff}\sim 8.8 a_0$. The increase of $\xi_{eff}$ by 
increasing $R_y$ is due to the enhancement of the inverse proximity 
effect which reduces the average gap in the S region.

On the other hand, panel (b) on Fig.~\ref{Rx_dependence}
illustrates the increase of $\xi_{eff}$ when a magnetic field is applied.
The coherence length for $\phi/\phi_0=0.6$ is $20\,\%$
larger than for $\phi/\phi_0=0$, which again can be associated to the
reduction of the average gap, now due to the 
depairing effect of the applied magnetic field.

As can be more clearly observed for low transparency, 
the transmission coefficients show small fluctuations
around $R_x\simeq p R_y$, with $p$ being an integer. This is another
feature associated with the ballistic character of our model
as it is  discussed in the  Appendix.

Finally, Fig.~\ref{fig:dep_flux_mg} shows the flux dependence of the
EC and CAR transmission coefficients. We find that this dependence
can be fitted as 
\begin{eqnarray}
T_{EC,CAR}(\phi)&=&T_{EC,CAR}(\phi=0)
\left[1+\left(\frac{\phi}{\phi_*}\right)^2\right] ,
\label{eq:fluxdependence}
\end{eqnarray}
where $\phi_*/\phi_0$ is roughly of order unity. 
For instance, the same value $\phi_*/\phi_0=1.35 \pm 0.05$ 
for the EC and CAR transmission coefficient
is obtained within error-bars for the data on 
Fig.~\ref{fig:dep_flux_mg}.

\subsection{Non-local conductance at arbitrary bias}

At arbitrarily large bias the non-local conductance cannot be
obtained from Eq. (\ref{eq:Gab1}) with 
the transmission coefficients defined by Eqs. 
(\ref{eq:TEC}) and (\ref{eq:TCAR}) as it contains contributions
due to the voltage dependence of the self-consistent order parameter
in the S region. More generally it can be computed directly as 
$\partial I_a/\partial V_b$.

The upper panels of Fig. \ref{fig:dep_voltage} illustrate
the behavior of ${\cal G}_{a,b}$ at arbitrary bias for 
a system size $(R_x,R_y)=(14 a_0,20 a_0)$ and two different values of the
normal transparency $T_N$. For small $T_N$ the non-local
conductance exhibits a abrupt jump for $e V_b \sim \Delta_0$.
In this limit ${\cal G}_{a,b}$ is well described by Eq. (\ref{eq:Gab1}) 
[see inset in panel (a)]
which is a consequence of having a quasi-equilibrium situation
in the whole voltage range. This is also reflected in the
magnetic field dependence which remains similar to the one
found at low bias. 
In contrast, for high transparency non-equilibrium effects
manifest in several features of the non-local conductance.
For instance one can clearly notice that the jump associated
to the gap appears at lower voltage bias (Fig.\ref{fig:dep_voltage} b). Also, as shown in the
inset, the actual value of ${\cal G}_{a,b}$ deviates from the
one calculated from Eq. (\ref{eq:Gab1}). Finally,  
it is found that the magnetic field at voltages $\sim \Delta_0$
reduces the non-local conductance, i.e. it
has the opposite effect to the one found at low bias.

In order to make contact with existing experiments it is also
interesting to analyze the behavior of the non-local resistance,
defined as
\begin{equation}
{\cal R}_{a,b} = -{\cal G}_{a,b}/[{\cal G}_{a,a} {\cal
    G}_{b,b} - {\cal G}_{a,b} {\cal G}_{b,a}]
\label{eq:res-nonlocal}
\end{equation}
where
${\cal G}_{a,a}$ and ${\cal G}_{b,b}$ denote the local
conductances at each interface and ${\cal G}_{a,b}$ and ${\cal G}_{b,a}$
are the non local conductances. The later take approximately the
same value under quasi-equilibrium conditions, but they are in general
different due to the voltage dependence of the
superconducting order parameter. Eq.~(\ref{eq:res-nonlocal}) is valid
provided that the induced voltage $V_a$ is very small
and under the assumption that the current voltage relations can
be linearized around the voltage $V_b$ under consideration.
The corresponding results are
shown in the lower panels of Fig. \ref{fig:dep_voltage}. 
While ${\cal R}_{a,b}$ is almost constant in the region of small voltages
it exhibits a high increase around the self-consistent gap.

Even stronger non-equilibrium effects are found when increasing
the transverse dimension $R_y$ and for higher values of the
normal transparency $T_N$ at the $b$ side while maintaining
at $a$ at low values. In that case most of the injected current
flows into the superconducting electrode and one can obtain an abrupt
jump in ${\cal R}_{a,b}$ from positive to negative for $V_b$ of the
order of the self-consistent gap \cite{note}.
The origin of this change of sign can be understood 
from the results in Fig. \ref{fig:desI} where the local and non-local
conductances are plotted for  different values of $R_x$
(upper panels). As can be observed, the local conductances
exhibit a decrease for voltages approaching the self-consistent
gap. This decrease can even lead to a negative local differential 
conductance
which arises due to the fact that the injected current can
become larger than the maximum supercurrent that can leak into
the superconducting electrode. A similar effect was described
in Ref.~\onlinecite{zaitsev}. This effect is reduced by increasing the length
$R_x$.

As a consequence of the suppression of the local conductances 
there exists a voltage window where
one can have ${\cal G}_{a,b}{\cal G}_{b,a} > {\cal G}_{a,a} {\cal G}_{b,b}$
which, according to Eq (\ref{eq:res-nonlocal}), leads to a change 
of sign in the non-local resistance.
It is worth emphasizing that this change of sign is not
related to a change of sign of ${\cal G}_{a,b}$ and therefore
not associated to a possible dominance of CAR over EC processes.
These results are in qualitative agreement with the experimental data of
Ref.~\onlinecite{Cadden}. 

We plot in the lower panels of Fig.~\ref{fig:desI} the variation of
${\cal R}_{a,b}$ as a function of $V_b$.
One can observe that the abrupt change of sign in ${\cal R}_{a,b}$ reduces its
amplitude and shifts towards higher voltages as $R_x$ increases. The 
shift corresponds to the increase of the critical current of the central
region which grows linearly with $R_x$, an effect which is not present
in the data of Ref.~\onlinecite{Cadden} as it corresponds to a 
somewhat different geometry.
Moreover, while our calculations have been performed in the voltage 
biased case, the experiments of Ref.~\onlinecite{Cadden}
correspond to a current biased 
situation. Thus, the regions of negative differential conductance could 
not be explored. However, the regions with negative ${\cal R}_{a,b}$ shown in 
the lower panels of Fig.~\ref{fig:desI} 
would be observable, since the sign change 
of ${\cal R}_{a,b}$ 
is related to the condition ${\cal G}_{a,a} {\cal G}_{b,b} < 
{\cal G}_{a,b} {\cal G}_{b,a}$ 
rather than to the change of sign of ${\cal G}_{b,b}$.
In fact, we have checked 
that the change of sign of ${\cal R}_{a,b}$
takes place outside the regions with 
negative ${\cal G}_{b,b}$.
%%%%%%%%%%%%%%%%%%%%%%%%%%%%%%%%%%%%%%%%%%%%%%%%%%%%%%%%%%%%%%%%%%%%%%
\begin{figure}[t]
\centerline{\includegraphics[scale=0.9]{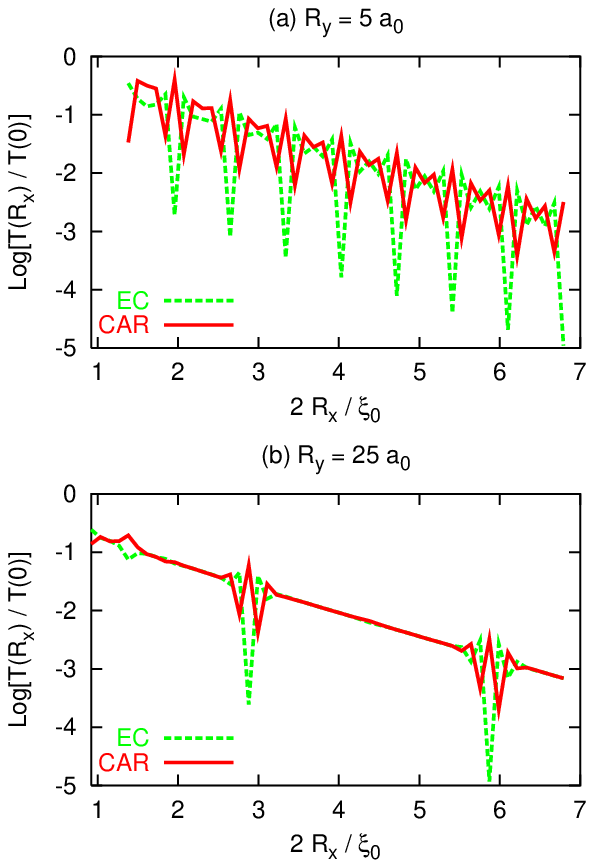}}
\caption{(Color online). 
Dependence of the normalized EC and CAR transmission coefficients 
$T(R_x)/T(0)$ on
$2 R_x/\xi_0$, where $R_x$ is the dimension of the superconductor
along $x$ axis, and $\xi_0$ is the bulk coherence length. 
The applied voltage is small compared to the gap ($e V_b/\Delta_0=0.02$).
The coherence length is larger than for the corresponding parameters
that in the presence of superconducting leads.
Panel (a) corresponds to $R_y=5\,a_0$
and panel (b) to $R_y=25\,a_0$.
The symbols correspond to EC and the bold red lines to CAR.
\label{fig:simplified_model}
}
\end{figure}
%%%%%%%%%%%%%%%%%%%%%%%%%%%%%%%%%%%%%%%%%%%%%%%%%%%%%%%%%%%%%%%%%%%%%%

\section{Conclusions}
\label{sec:conclu}

We have presented self-consistent model calculations for describing
a typical experimental setup to measure non-local transport in 
nanoscale superconductors connected to normal electrodes.
An important ingredient in these 
calculations is the inclusion of the superconducting leads
which drain the current injected into the nanoscale region.

We have found that this system exhibits a rich variety of
behaviors controlled by several parameters. One can  distinguish
two limiting cases: the quasi-equilibrium regime in which
the superconducting order parameter depends only weakly on
the applied voltage and the strong non-equilibrium case where
this dependence cannot be neglected. In these two regimes
the non-local conductance exhibits a quite different behavior.
In quasi-equilibrium conditions ${\cal G}_{a,b}$ can be
expressed as the difference between $T_{CAR}$ and $T_{EC}$.
We have shown that $T_{EC}$ becomes larger than
$T_{CAR}$ for increasing transparency, and that both decay
exponentially with the system size $R_x$. Moreover we have
shown that both coefficients increase under an applied magnetic
field. 

In the case of strong non-equilibrium a simple expression of
${\cal G}_{a,b}$ as the difference between $T_{CAR}$ and $T_{EC}$
is no longer possible. The voltage dependence of the self-consistent
order parameter introduces an additional contribution to the
non-local conductance. We have also found that even the local 
conductance can be strongly modified by non-equilibrium effects.
The effects on these two quantities lead to a non-local resistance
which may exhibit a strongly non-monotonous behavior as a function 
of the injected current including a sign change for sufficiently
large transparency. 

As a final remark we would like to comment the possible connection 
between the present model calculations and the existing experiments on
non-local transport. As mentioned in the introduction,  basically
three different experiments have been performed so far. 
They correspond to different geometries and to different materials.
The size of the junctions transparency can also change enormously from
one experiment to the other. It is thus not possible to infer general
conclusions from these results. In the case of the experiment
of Ref.~\onlinecite{Russo} the authors claim to have junctions with a extremely
small transparency $\sim 10^{-5}$ which would warrant the quasi-equilibrium
conditions in the subgap voltage range. The results of the present work,
corresponding to a non-interacting theory, would not be able to describe
that experiment. In particular our results predict an increase of the 
non-local conductance with magnetic field in contradiction with the
experimental observations. As stated in a previous work by some of
the authors \cite{Levy-Yeyati}, interactions mediated by the electromagnetic 
modes may play an important role in that experimental situation.
It was also conjectured \cite{Duhot-Melin-PRB-Hikami}
that disorder may play a role when associated to phase fluctuations.

On the other hand, the experimental results of Ref.~\onlinecite{Cadden} clearly
correspond to the strong non-equilibrium situation described in the
present work. Even when our model geometry does not fully correspond
the their experimental setup the behavior of the non-local resistance
as a function of the injected current shown in Fig.\ref{fig:desI}
is very similar to the one found in this experiment. 
A closer comparison between theory and experiments would be
desirable for further understanding of the observed features.  

\section*{Acknowledgements}
The authors thank D. Feinberg for continuous interest in non
local transport, and S. Florens for useful reading of the manuscript.
The authors acknowledge support by Spanish MCYT through projects 
FIS2005-06255 and NAN2007-29366-E. 
R. M. acknowledges support from the Agence Nationale de la Recherche
of the French Ministry of Research under contract Elec-EPR.
F. S. B. acknowledges funding by the Ram\'on y Cajal program.

\bigskip

\appendix*
\section{Green functions and transmission coefficients of a closed rectangular superconducting cavity}
In order to understand the 
oscillations in the
EC and CAR transmission coefficients as a function of the length $R_x$
(Fig.\ref{Rx_dependence}),
we consider in this Appendix a simplified model consisting of
a central superconducting region which we describe by 
the tight-binding Hamiltonian of Eq.~(\ref{eq:H-BCS})
on a sample of dimensions $R_x\times R_y$, with
$R_x=N a_0$ and $R_y = M a_0$. We do not include the contacts
to  the superconducting  leads and do not implement self-consistency for the
order parameter.

We assume tunnel contacts with the normal electrodes at left and right so
that the non-local conductance is given by lowest order
perturbation theory in the tunnel amplitudes.
The  normal Green function describing a particle propagating from  site $\alpha$ at coordinates $(x_\alpha,y_\alpha)$
to site $\beta$ at coordinates $(x_\beta,y_\beta)$ is given by\cite{Melin-EPJB}:
\begin{widetext}
\begin{eqnarray}
\label{eq:g11}
\left[g_{\alpha,\beta}(\omega)\right]_{11} &=&
\left(\frac{2}{1+N}\right)
\left(\frac{2}{1+M}\right)
\sum_{n=1}^{N} \sum_{m=1}^M
\sin{\left(n\frac{x_\alpha/a_0+1}{N+1}\right)}
\sin{\left(n\frac{x_\beta/a_0+1}{N+1}\right)}\nonumber\\
&&
\label{eq:g11bis}
\sin{\left(m\frac{y_\alpha/a_0+1}{M+1}\right)}
\sin{\left(m\frac{y_\beta/a_0+1}{M+1}\right)}
\left[\frac{\omega+i\eta+\xi_{n,m}}
{(\omega+i\eta)^2-\Delta^2-(\xi_{n,m})^2}\right]
,
\end{eqnarray}
\end{widetext}
where the quasiparticle energy $E_{n,m}$ is such that
$E_{n,m}^2=\Delta^2+(\xi_{n,m})^2$, with
$\xi_{n,m}=2t [\cos{(\frac{n\pi}{1+N})}+
\cos{(\frac{m\pi}{1+M})}]$. The coherence factors are given by
$(u_{n,m})^2=(1/2)(1+\xi_{n,m}/E_{n,m})$,
$(v_{n,m})^2=(1/2)(1-\xi_{n,m}/E_{n,m})$.
The anomalous component $\left[g_{\alpha,\beta}\right]_{12}$
is obtained by substituting the numerator $\omega+i\eta+\xi_{n,m}$
in Eq.~(\ref{eq:g11bis}) by $-\Delta$.
These Green functions lead to an enhanced probability for the propagation
along certain directions. As one would expect from a semi-classical
analysis of an integrable cavity these directions correspond to
the diagonal lines in the case of a square superconducting region.
The ripples shown by the self-consistent gap are a consequence of 
constructive interference along these semi-classical trajectories.

The EC and CAR transmission coefficients can be evaluated
using these approximated Green functions.
Fig.~\ref{fig:simplified_model} shows the behavior of 
the transmission coefficients $T_{EC}$
and $T_{CAR}$ as a function
of $R_x$ obtained with this simplified model
for two values of $R_y$.  We find that the the EC and CAR transmission 
coefficients almost coincide for $R_x$ in between $p R_y$ and $(p+1)R_y$, 
where $p$ is an integer. Pronounced oscillations are obtained when the 
dimensions of the superconducting region are such that $R_x/R_y\simeq p$.  
Again these effects are a consequence of interferences along semi-classical 
trajectories in this ballistic integrable model.
In the same way, one could use these approximate Green functions
in a self-consistent calculation to show that the ripples in the
self-consistent gap (see e.g. Fig.2\ref{fig:des_gap}) have a similar origin.


\begin{thebibliography}{99}
\bibitem{note} We found no abrupt jump for a symmetric junction
for $T_N=1$ and with sample
dimensions $R_x=12 a_0$ and $R_y=20 a_0$
corresponding to Fig.~\ref{fig:dep_voltage}.
\end{thebibliography}
\end{document}